\documentclass[5p]{elsarticle}
\usepackage{graphicx}

\title{Obtaining the diffusion coefficient for cosmic ray propagation in the Galactic Centre Ridge through time-dependent simulations of their $\gamma$-ray emission}

\author[astr]{Stavros Dimitrakoudis\corref{cor1}}
\ead{sdimis@phys.uoa.gr}

\author[astr]{Apostolos Mastichiadis}
\author[nuc]{Athanasios Geranios}

\cortext[cor1]{Corresponding author, sdimis@phys.uoa.gr}

\address[astr]{University of Athens, Physics Department, Section of Astrophysics, Astronomy and Mechanics, Panepistimioupoli 15771, Greece}

\address[nuc]{University of Athens, Nuclear and Particle Physics Department, Panepistimioupoli 15771, Greece}

\begin{document}

\begin{abstract}

Recent observations by the H.E.S.S. collaboration of the Galactic Centre region have revealed what 
appears to be $\gamma$-ray emission from the decay of pions produced by interactions of recently 
accelerated cosmic rays with local molecular hydrogen clouds. Synthesizing a 3-D hydrogen cloud map from 
the available data and assuming a diffusion coefficient of the form $\kappa(E) = 
\kappa_0(E/E_0)^\delta$, we performed Monte Carlo simulations of cosmic ray diffusion for various 
propagation times and values of $\kappa_0$ and $\delta$. By fitting the model $\gamma$-ray spectra to 
the observed one we were able to infer the value of the diffusion coefficient in that environment 
($\kappa = 3.0 \pm 0.2 kpc^2Myr^{-1}$ for $E = 10^{12.5}eV$ and for total propagation time $10^4 yr$) as 
well as the source spectrum ($2.1 \leq \gamma \leq 2.3$). Also, we found that proton losses can be 
substantial, which justifies our approach to the problem.

\end{abstract}

\maketitle

\section{Introduction}

The High Energy Stereoscopic System (H.E.S.S.) $\gamma$-ray telescope began operations in 2004 and
among its first targets was the Galactic Centre region \cite{Hinton2006}. 
Images obtained from the H.E.S.S. observations had an angular resolution $0.1^\circ$, giving us a map of
 $\gamma$-ray emission of unprecedented detail. A few point sources were first detected, which were
compatible with the positions of known objects, like the black hole Sagittarius A*, the supernova
remnant Sgr A East and the supernova remnant/pulsar wind nebula G0.9+0.1. By subtracting those known
sources, the H.E.S.S. collaboration was able to produce a mapping of fainter $\gamma$-ray emission, stretching
over an area of approximately 300pc by 100pc around the Galactic Centre \cite{Aharonianetal2006}. This
emission seems to follow the contours of molecular gas density, as measured by its carbon monosulfide
distribution \cite{Tsuboietal1999}, up to a distance of approximately $1.3^\circ$ in galactic longitude from
the Galactic Centre. Beyond that distance the $\gamma$-ray emission diminishes to background levels. 

This apparent emission correlation points towards possible models, for which two theories were
initially proposed \cite{Aharonianetal2006}. The first postulated that a population of electron
accelerators produces the observed emission via inverse Compton scattering. The objects that would make
up such a population, such as pulsar wind nebulae, would thrive in regions of high-density molecular
gas; however the large number of sources needed to reproduce the observed emission renders this
possibility rather unlikely \cite{Aharonianetal2006}. 

The other theory claims that the $\gamma$-rays are produced by neutral pion decay, resulting from
the interaction of locally accelerated cosmic rays with the ambient molecular gas. The lower energy
threshold of the H.E.S.S. survey was 380 GeV, so cosmic ray protons of higher energies would be needed
to produce the observed $\gamma$-rays. Furthermore, these cosmic rays would have to have been
accelerated near the Galactic Centre at some point in the past, yet not diffused significantly beyond
$1.3^\circ$ from it. Assuming the validity of this theory, one may use this data to infer the diffusion
properties of cosmic rays in the Galactic Centre region for various possible sources of cosmic rays.
Considering as a source of the cosmic rays either the supernova remnant Sgr A East, with an estimated
age of 10 kyr \cite{Uchidaetal1998}, or the black hole Sgr A* with a more remote age of activity,
Aharonian et al. \cite{Aharonianetal2006} inferred that the diffusion coefficient should be no more
than  $3.5 kpc^2 Myr^{-1}$ in that area. Busching et al. \cite{Buschingetal2006} went one step further
with an analytical reproduction of the H.E.S.S. observations by calculating the emission for different
diffusion coefficients. Using relativistic protons of mean energy $\sim 3 TeV$ to represent the cosmic
rays  and a small number of Gaussian functions to represent the molecular clouds they arrived at a
diffusion coefficient of  $\kappa = 1.3 kpc^2 Myr^{-1}$. B\"usching and de Jager 
\cite{BuschingJager2008}
subsequently expanded their results for different ages and on-times for the source. More recently,
Wommer at al. \cite{Wommeretal2008} employed a different approach, in which the motion of individual
test particles is computed by solving the Lorentz force equation for short time periods, and the
resulting distributions provide the coefficients for the diffusion equation. From their initial
conditions for the turbulent magnetic field, it is derived that no single source can account for the
observed emission, but that a more continuous source of cosmic ray protons produces a better
correlation.

In this paper we present a series of time-dependent simulations of proton propagation in the Galactic
Centre in a detailed 3D distribution of molecular hydrogen concentrations for different diffusion 
parameters. By utilizing a wide array of data on molecular clouds in the Galactic Centre from various 
observations we construct a fairly accurate density grid of the area of diffusion, as shown in \S 2.1. 
We can then inject cosmic rays, assumed to be protons, from an origin of our choice, which propagate 
according to the diffusion model and its free parameters described in \S 2.2. In \S 2.3 we deal with the 
interactions of these protons with hydrogen molecules and with the production of photons from the 
resulting neutral pion decays. In \S 2.4 we briefly discuss the likely sources of cosmic ray origin, and 
the implications on their age. In \S 2.5 we describe the more technical aspects of the simulation 
program, such as the initial choice of free parameters and the methods used to raise the efficiency of 
the numerical code. Finally, in \S 3 we present the results of the simulations and in \S 4 we summarize 
and discuss the effectiveness of our approach and the likely significance of our results. The present 
paper expands upon the initial results of Dimitrakoudis et al. \cite{Dimitrakoudis2008a}.

\section{Simulations}

\subsection{Synthesizing a hydrogen cloud map }

The area of the Galactic Centre in which we simulated the diffusion of cosmic rays is rich in $H_2$ 
gas, contained in a complex setup of high-density clouds, ridges and streams that comprise about 10\% of 
our galaxy's interstellar molecular gas, i.e. about 2 to 5 $\times 10^7$ solar masses 
\cite{Tsuboietal1999}. Due to the high densities involved, tracer molecules were used to determine the 
mass of each gas cloud. To create a realistic 3-D map of that environment we first obtained the data for 
the 159 distinct molecular cloud clumps, as compiled by Miyazaki \& Tsuboi \cite{MiyazakiTsuboi2000}, 
i.e. we assigned to each one a galactic longitude, latitude, radius and density. Their radial distances 
are unknown, so we assumed a random function to simulate them, within reasonable limits. To these data 
we added the locations and densities of the radio-sources Sgr A \cite{Shuklaetal2004}, Sgr B1, Sgr B2 
and Sgr C \cite{LawYusefZadeth2004}, which are rich in atomic hydrogen. We then broke down the entire CS 
map of the Galactic Centre by Tsuboi et al. \cite{Tsuboietal1999} into blocks of uniform density, which 
we treated as clouds of equal radius and of random radial distances. Finally we added the larger CO 
clouds by Oka et al. \cite{Okaetal1998} in order to expand the spatial distribution of our map. Every 
time we added a new set of clouds we checked the possibility that clouds from previous sets were sharing 
their projected locations with the new ones, and we subtracted their masses accordingly. The result was 
584 clouds of hydrogen gas with a high uncertainty as to their radial distances. These clouds were then 
turned into a 3-D grid of $120 \times 60 \times 60$ boxes of uniform density, which form the volume of 
our diffusion model, $5.4 \cdot 10^7 pc^3$. The projection of that grid constitutes an area far greater 
than the extent of the observed ã-ray emission by \cite{Aharonianetal2006}, comprising a total mass of 
$1.2 \cdot 10^8 M_\odot$, so particle interactions with clouds well outside the observed area of 
emission are accounted for.
   
\subsection{Diffusion model }

We used the diffusive model of CR propagation, assuming a diffusion coefficient of the form: 
$$\kappa = \kappa_0(E/E_0)^\delta$$
where E is the energy of the CR protons, $E_0=1GeV$, while $\kappa_0$ and $\delta$ are free parameters, 
whose values we will try to infer through our simulations. The index $\delta$ is a measure of the 
turbulence of the magnetic field and is assumed to be $0.3 \leq \delta \leq 0.6$ \cite{Strongetal2007}, 
while $\kappa_0$ is a constant whose value we will attempt to estimate once $\delta$ has been obtained. 
The diffusion coefficient in our simulations is represented by the mean free path $\ell$, which is given 
by the usual expression 
$$\ell = 3\kappa/c$$
where c is the velocity of light.

Thus, the test protons move in straight lines of length equal to $\ell$, after which their directions 
change randomly. At the end of each such free walk, the box number of the hydrogen density grid is 
calculated and a check is made for collisions with hydrogen protons. If such a collision occurs, the 
diffusion continues with a lower proton energy, as shown in the next section. 

\subsection{Production of $\gamma$-rays }

We have assumed that all $\gamma$-rays forming the observed emission are produced from neutral pion 
decay. Moreover, we have assumed that pions are produced in proton-proton collisions through two main 
channels

\
   
a) $p + p \rightarrow p + p + \pi^+ + \pi^- + \pi^0$

\

b) $p + p \rightarrow n + p + \pi^+$

\

In case (a) the initial proton loses part of its energy and a multiplicity of pions is created that take 
equal amount of the energy lost from the proton. The positive and negative pions break down into muons, 
and ultimately into electrons, positrons and neutrinos, while the neutral pions decay into $\gamma$-
rays. Assuming an inelasticity $k_{pp}=0.45$ \cite{MastichiadisKirk1995}, 15\% of the initial proton 
energy will go to the produced $\gamma$-rays, while the original proton will continue its diffusion with 
55\% of its initial energy. This approach, while simplified, gives results which are in good agreement 
with the more detailed spectra produced by B\"usching et al. \cite{Buschingetal2006}.

Channel (b) produces no photons and thus does not contribute directly to our observed emission. However, 
the initial proton is turned into a neutron, having lost some of its energy in the collision, and will 
thus continue its propagation in a straight line, unaffected by the magnetic fields that regulated its 
trajectory as a proton. It will revert, though, back into a proton after half life $\tau = 886.7 \pm 0.8 
sec$ \cite{Yaoetal2006}, and then it will continue its diffusion as before. Accounting for time 
dilation, the distances traveled by neutrons are always much smaller than the mean free path for each 
proton energy (0.028pc for $E=10^{12.5}eV$; 0.92pc for $E=10^{14}eV$; the radii of molecular gas clouds 
range between 2pc and 40pc), so we can safely assume that this case will not affect the diffusion 
process.

Using the above we calculate the optical depth for each random walk. This is calculated as the product 
of the interaction cross section with the mean free path $\ell$ and the density n of hydrogen molecules. 
That density is retrieved from the grid box where each step ends, a fairly effective approximation as 
most random walks are contained within single grid boxes. We have assumed that the cross section is 
$\sigma_{pp} = 4 \cdot 10^{-26} cm^2$ \cite{Begelmanetal1990}, while an increase by a factor of 1.3 is 
sufficient to account for the known chemical composition of the Interstellar Medium 
\cite{MannheimSchlickeiser1994}. The reaction probability would then be $P = 1 - e^{-\tau}$, where 
$\tau$ is the optical depth. In practice though, the optical depth is a very good approximation of $P$ for 
low probability values ($P < 0.15$). Thus, to speed up our simulations, we can simply define the 
reaction probability as the optical depth, having first checked that its value is always low enough to 
warrant the approximation. At every step, that probability is checked against a random number. If the 
random number is smaller than the reaction probability, then there is a collision and $\gamma$-rays are 
produced, and the proton continues its propagation with diminished energy. In the case where its final 
energy is so low that we are no longer interested in its resulting photons, the proton is removed from 
the simulation.

Apart from proton-proton collisions, CR protons could lose energy due to ionization losses, coulomb 
collisions, Compton scattering, synchrotron emission, Bethe-Heitler pair production and photo-pion 
production. Wommer et al. have demonstrated that energy loss rates due to the four latter processes are 
insignificant compared to energy loss rates from proton-proton collisions \cite{Wommeretal2008}. As for 
ionization and Coulomb losses, Mannheim and Schlickeiser \cite{MannheimSchlickeiser1994} clearly show 
that their effect is negligible compared to that of proton-proton collisions in the local interstellar 
medium. In the higher density environment of the galactic centre, and especially within molecular 
hydrogen clouds, their comparative effect would be minimal.

\subsection{Possible sources}

Assuming a single source for the origin of the diffuse cosmic rays in the Galactic Centre, then the most 
likely candidates would be either the supernova remnant Sgr A East or the black hole Sgr A*. The former 
has an estimated age of $10^4 yr$ \cite{Uchidaetal1998} while the latter could have had a burst of 
activity even further in the past. We have conducted simulations for both scenarios, selecting an age of 
$10^4$ yr for the former case and an age of $10^5$ yr for the latter. In both cases, the galactic 
coordinates for the source are set to $l = 0^\circ$, $b = 0^\circ$, which correspond well to the 
location of Sgr A* and to the approximate centre of the extended source Sgr A East. 

\subsection{Simulation parameters}

Except for the time available for diffusion (which corresponds to the elapsed time since activity at the 
source), all other parameters are the same for both potential sources. We have assumed that the source 
is active for 100 years and that it produces a constant flux of cosmic rays during that period. During 
this time cosmic rays are assumed to constantly escape from the source and to start diffusing. The 
proton injection spectrum is divided into six logarithmic bins, ranging from $E = 10^{12.5} eV$ to $E = 
10^{15} eV$. Lower energies would  produce $\gamma$-rays below the H.E.S.S. threshold, while higher 
energies would have a negligible impact on the resulting emission, due to their low numerical density 
and their large mean free paths. Each bin has the same number of test particles, which is fixed for each 
run, depending on the interaction probabilities for a given diffusion coefficient. Those numbers are 
then weighted after each simulation according to a power law distribution that fits best the power law 
of the observed $\gamma$-rays from H.E.S.S., see \cite{Aharonianetal2006} ($\Gamma = 2.29 \pm 0.07_{stat} 
\pm 0.20_{sys}$). Since the weighting procedure is applied to the photon spectrum at the end of the 
simulation, the resulting initial proton spectrum is generated in deference to its modification during 
the diffusion process.

To improve the efficiency of the simulations, we treat the process of continuous production of cosmic 
rays in the following way. Only one burst of cosmic rays of all energies is created in the simulation, 
starting at the beginning of the source's activity. The photons produced at the end of the simulation 
are, naturally, the ones observed from protons that have propagated for the duration of our simulation. 
In addition to them we also take into account all the photons produced in a time period before the end 
of the simulation that is equal to the production time at the source ($10^2 yr$). These correspond to 
the photons that would have been produced at the end of our simulation by newer populations of protons 
produced within that time range at their source. The actual production time is of little importance, as 
long as it much shorter than the propagation time (as is also seen in \cite{BuschingJager2008}). Had we 
considered all the protons to have been produced instantaneously, the results would have been the same, 
provided the number of protons was increased to provide the same statistical robustness. On the other 
end, a production time as large as $10^3 yr$ wouldn't significantly alter our results. 

Other parameters besides the two ages are the index $\delta$ and the proportionality factor $\kappa_0$ 
of the diffusion expression. For the index $\delta$ we have chosen the values 0.3, 0.4, 0.5 and 0.6, 
while $\kappa_0$ takes on a series of test values by increments of log(0.1) around the roughly expected 
value required for a diffusion coefficient that would allow a mean propagation of $<r>^2 = \kappa \cdot 
\Delta t$, in accordance with each total time and the mean propagation distance inferred from the 
observations. 

Once each simulation is completed and the results normalised, the resulting $\gamma$-rays are compared 
against the results from H.E.S.S. using the reduced $\chi^2$ criterion.

\section{Results}

The resulting reduced $\chi^2$ values for cosmic rays originating from the supernova remnant Sgr A East, 
assumed to be produced $10^4 yr$ ago, are shown in Fig.1 for various values of the diffusion 
coefficient. The minimum value of $\chi^2$ calculated is 1.8, and corresponds to $\delta = 0.3$ and 
$\kappa_0 = 0.25 kpc^2Myr^{-1}$. However we can find minima which are less than 2 for all values of 
$\delta$. If we use those values of $\delta$ and $\kappa_0$ to calculate the diffusion coefficient for 
protons of energy $10^{12.5}eV$ (the lowest energy in our sample and also the most important due to 
their relative abundance over those at higher energies), we arrive at the results illustrated in Fig.2. 
We can see that, for each value of $\delta$, the diffusion coefficient displays the same minimum at 
$\kappa = 3.0 \pm 0.2 kpc^2Myr^{-1}$. This value is higher than that calculated by B\"usching et al. 
\cite{Buschingetal2006} but close to that suggested by Aharonian et al. \cite{Aharonianetal2006} (less 
than $3.5 kpc^2 Myr^{-1}$).
   
\begin{figure}
\includegraphics [width=0.48\textwidth]{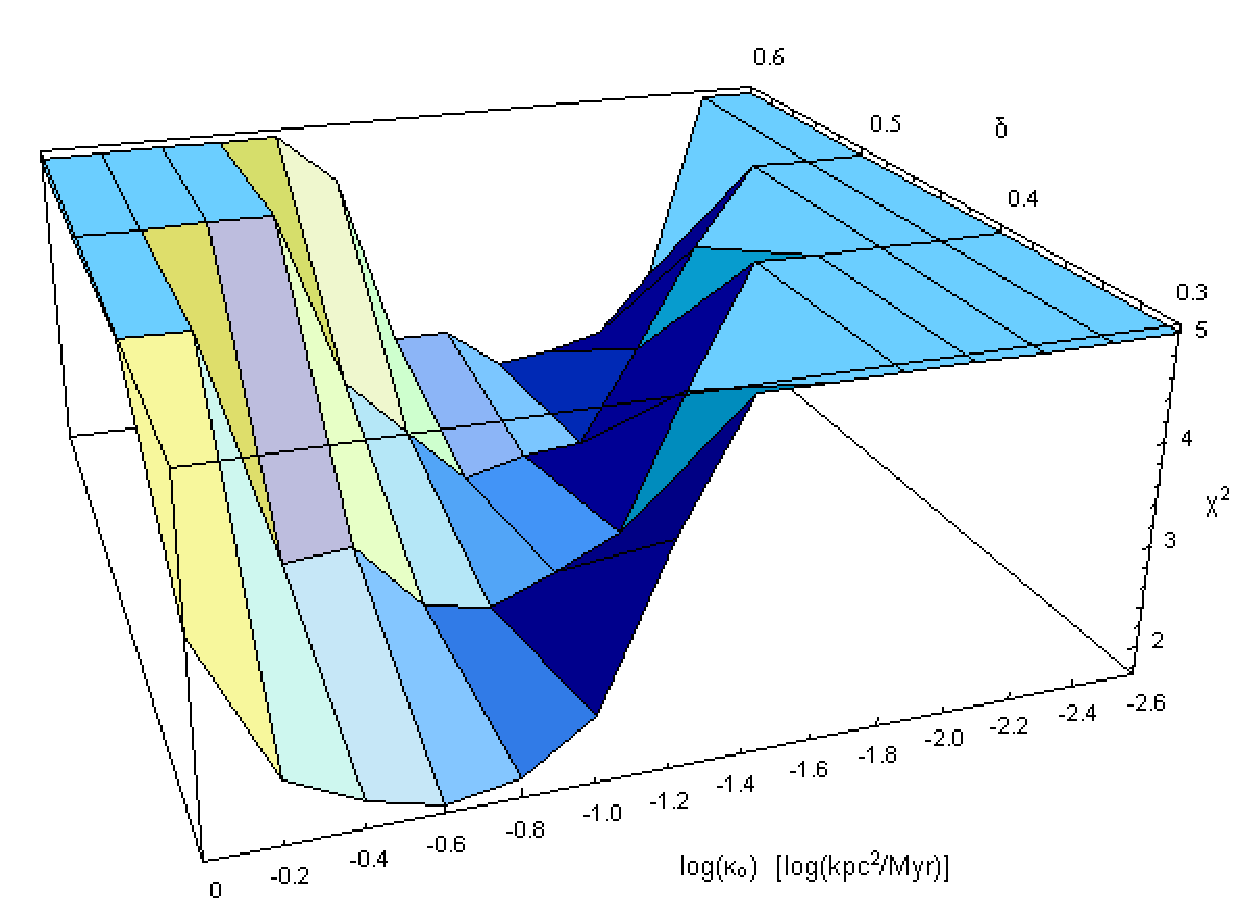}
\caption{Reduced $\chi^2$ values for different values of $\kappa_0$ and $\delta$ for total propagation 
time $10^4 yr$.}\label{fig1}
\end{figure} 

\begin{figure}
\includegraphics [width=0.48\textwidth]{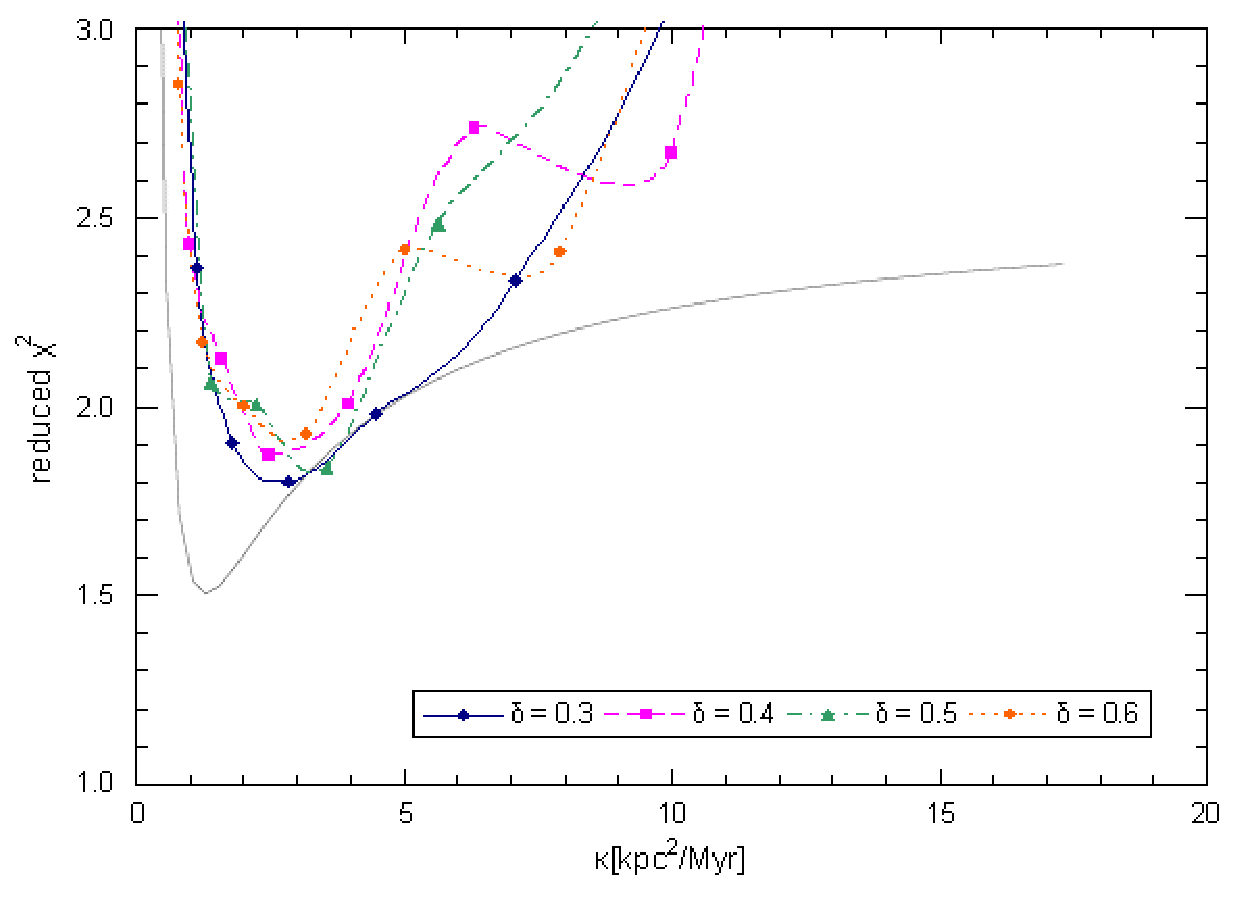}
\caption{The curves represent the reduced $\chi^2$ values for different diffusion coefficients, 
calculated for protons of energy $10^{12.5}eV$, for each value of $\delta$. We see in all cases a 
minimum for $\kappa \approx 3 kpc^2 Myr^{-1}$. The grey line shows the equivalent results from B\"usching 
et al. [3] for comparison.}\label{fig2}
\end{figure}  
   
For higher values of $\kappa$, we also see a disparity between the way the reduced $\chi^2$ values 
increase in our results and those of B\"usching et al. The reason for this is that for these values the 
mean free path becomes very large and comparable to the dimensions of the area of simulation, therefore 
the simulation becomes inefficient, resulting in poor statistics even with increased numbers of test 
particles. 
   
Fig.3 shows the reduced $\chi^2$ values that correspond to a total propagation time of $10^5 yr$ – a 
possible past activity of Sgr A*. If we once again use the values of $\delta$ and $\kappa_0$ which 
correspond to a minimum to calculate the diffusion coefficient for protons of energy $10^{12.5}eV$ we 
arrive at the results illustrated in Fig. 4. The resulting best choices for the diffusion coefficients 
appear in Table 1. 

\begin{table}[where]
\centering 
\begin{tabular}{c c} 

$\delta$ & $\kappa [kpc^2Myr^{-1}]$ \\
\hline 
0.3 & 0.45  \\ 
0.4 & 0.4  \\
0.5 & 0.56  \\
0.6 & 0.32  \\ 
\hline 
\end{tabular}
\caption{Diffusion coefficients for different values of $\delta$ for total propagation time $10^5 yr$.}
\end{table}
   
This variation for different values of $\delta$ is more pronounced than in the case of propagation for 
$10^4 yr$, but if we were to derive a mean value of $\kappa$ like before, we would find $\kappa = 0.43 
\pm 0.05 kpc^2Myr^{-1}$. The higher propagation time requires the cosmic rays to diffuse more slowly 
than in the case of $10^4 yrs$, which is why the resulting diffusion coefficient is considerably 
smaller.
   
\begin{figure}
\includegraphics [width=0.48\textwidth]{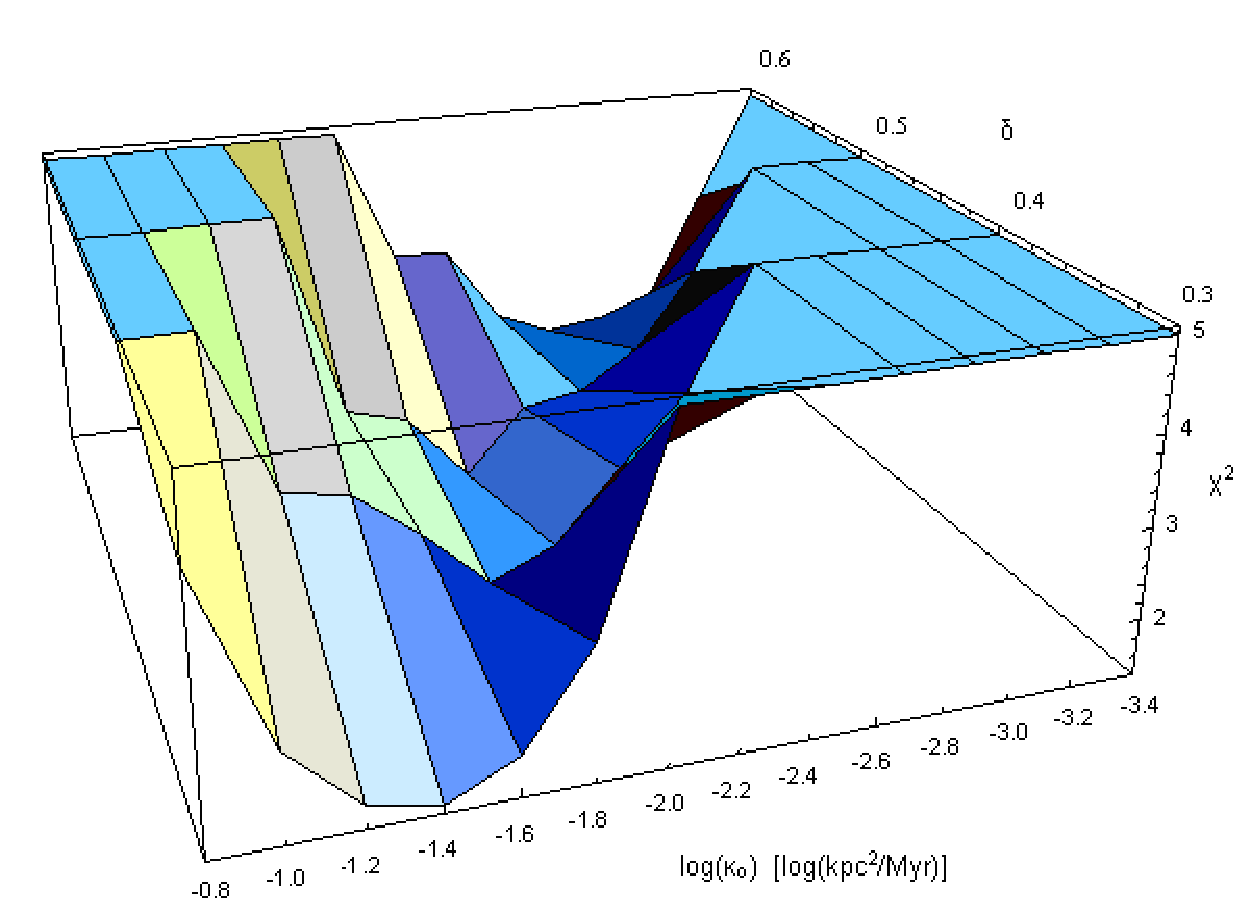}
\caption{Reduced $\chi^2$ values for different values of $\kappa_0$ and $\delta$ for total propagation 
time $10^5 yr$.}\label{fig3}
\end{figure} 

\begin{figure}
\includegraphics [width=0.48\textwidth]{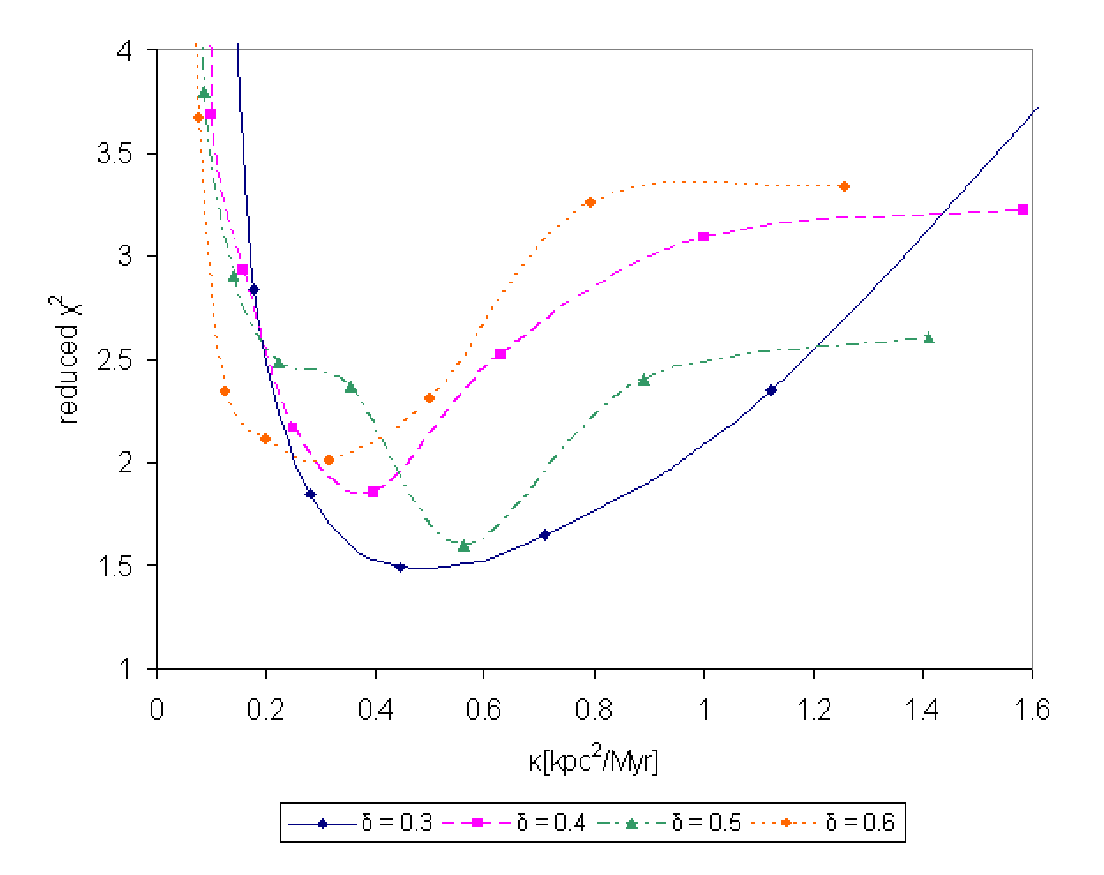}
\caption{Plot of the reduced $\chi^2$ as a function of the diffusion coefficient $\kappa$ for different 
values of $\delta$. The total propagation time was assumed to be $10^5 yr$. Higher values of $\delta$ 
tend to favor lower values of $\kappa$. }\label{fig4}
\end{figure}  

Fig.5 and 6 plot the proton spectral indices $\gamma$, as these were inferred from the simulations, for 
the two total propagation times versus $\kappa_0$ for different values of $\delta$.

\begin{figure}
\includegraphics [width=0.48\textwidth]{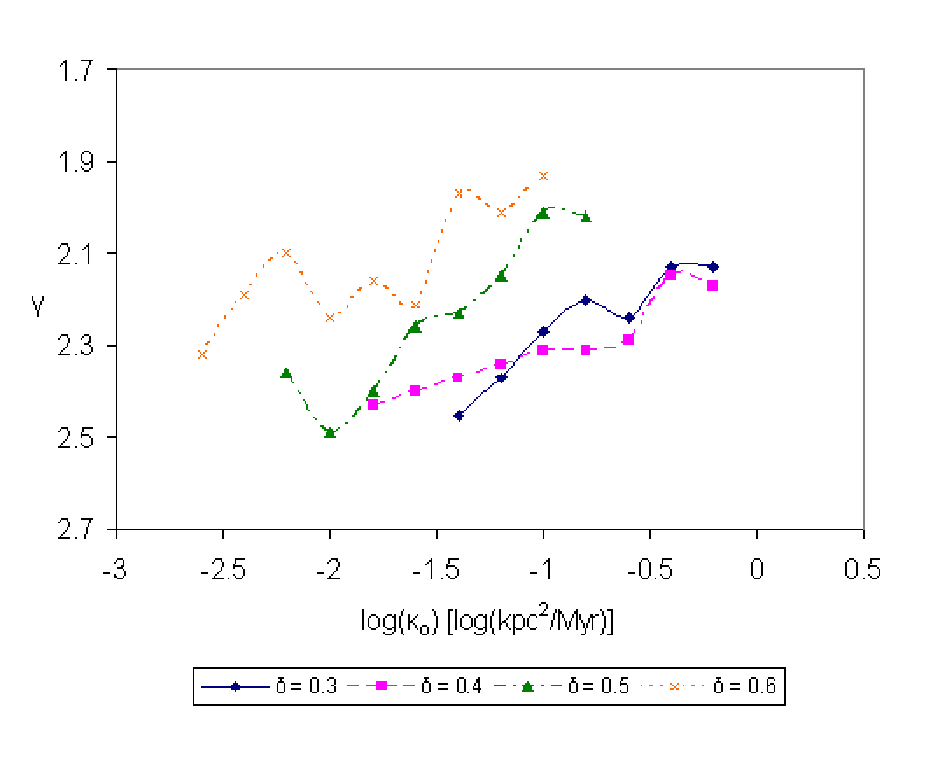}
\caption{Plot of spectral index $\gamma$ inferred from simulations as a function of the diffusion 
normalization $\kappa_0$ for different values of $\delta$. The total propagation time was taken to be 
$10^4 yr$.}\label{fig5}
\end{figure}

\begin{figure}
\includegraphics [width=0.48\textwidth]{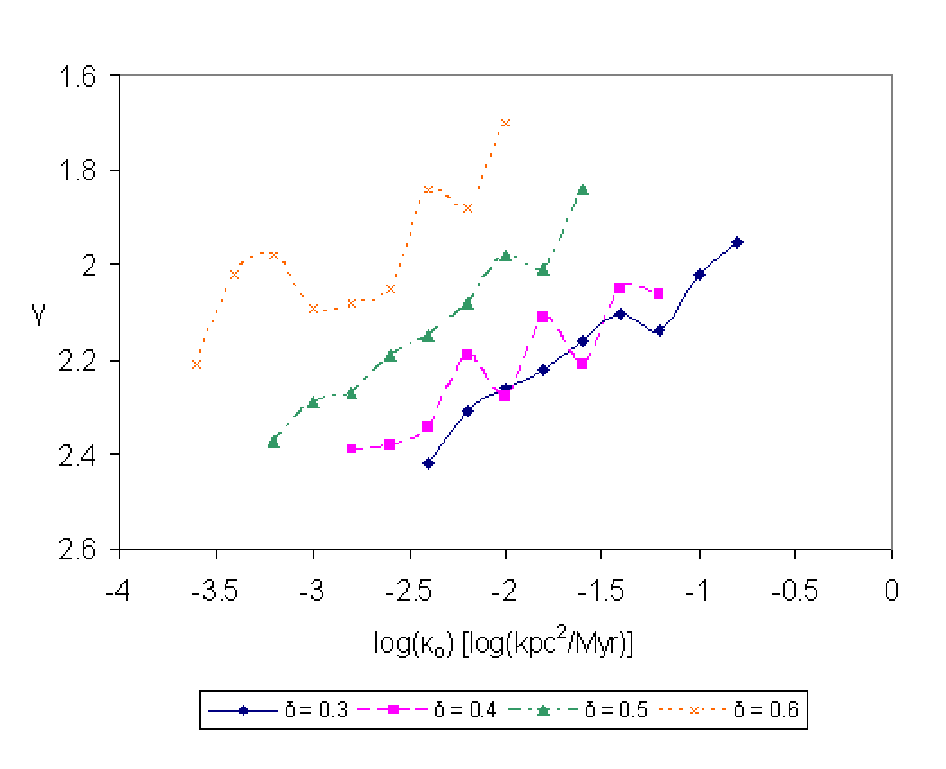}
\caption{Plot of spectral index $\gamma$ inferred from simulations as a function of the diffusion 
normalization $\kappa_0$ for different values of $\delta$. The total propagation time was taken to be 
$10^5 yr$. }\label{fig6}
\end{figure}
   
In Fig.5 we can see a fluctuation of $\gamma$ values, most likely due to irregularities in the 
interactions of the different energy populations with the complex setup of hydrogen clouds in the small 
propagation time. A progression to lower $\gamma$ values as we raise the value of $\kappa_0$ is, 
however, evident. The values of $\gamma$ for the minimum reduced $\chi^2$ values are in the range $2.1 
\leq \gamma \leq 2.3$. For the larger propagation time, we can see in Fig.6 a similar progression to 
lower $\gamma$ values as a function of $\kappa_0$. In the range of our best results, the proton indices 
are in the range $2.0 \leq \gamma \leq 2.1$, so the proton spectrum is clearly slightly steeper than the 
resulting photon spectrum. These results are in general agreement with leaky box model predictions 
\cite{Hillas2005}. 

Furthermore, it is possible to make an estimate of the total energy of the protons accelerated at their 
source, by comparing the number of observed photons to the number of diffusing protons. Doing so for the 
lowest observable proton energy and extrapolating for the energy range $10^9-10^{15} eV$ yields a total 
energy of $(8 \pm 1) \times 10^{49} erg$, for our different time and optimal diffusion parameters. That 
represents approximately 10\% of the energy output of a typical supernova explosion.

The Fermi Gamma-ray Space Telescope should be able to observe that area of space in a lower energy range 
(from 20 MeV up to 300 GeV). We have repeated our simulations for the optimal diffusion coefficients we 
have found, for a total propagation time $10^4 yr$, with the proton injection spectrum extended to $10^9 
eV$. In these cases, the emission is dominated by the lower energy protons, and thus extends from about 
$-0.3^\circ$ to $0.2^\circ$ in galactic longitude.

\section{Summary/Discussion}

We have presented the results of a series of time-dependent simulations of the diffusion of cosmic rays 
in the Galactic Centre region. In the first scenario it was assumed that a burst of cosmic rays occurred 
$10^4 yrs$ ago, while in the second $10^5 yrs$ ago. Likely candidates could have been the SNR Sgr A East 
or the black hole candidate Sgr A*. Their $\gamma$-ray emission was compared with observations from the 
H.E.S.S. collaboration \cite{Aharonianetal2006}, in order to determine the diffusion coefficient in that 
region. For that purpose, a detailed 3-D map of hydrogen concentrations
was synthesized from various observations \cite{LawYusefZadeth2004, MiyazakiTsuboi2000, Okaetal1998, 
Shuklaetal2004, Tsuboietal1999}, and two scenarios for the origin of the cosmic rays were taken into 
account. For the SNR Sgr A East, $10^4 yrs$ is the estimated upper limit on its age, so the resulting 
diffusion coefficient should be taken as a lower limit. There is much more uncertainty concerning the 
possible age of activity of Sgr A*, but it is clear that should that have been in the order of $10^5 
yrs$ ago or earlier, the random component of the interstellar magnetic field would have to be 
considerably more pronounced than in the first case.

The need for such elaborate simulations as were described in the preceding chapters arises from the 
considerable losses of protons during their propagation, losses which are inextricably connected to the 
local densities of hydrogen gas. In Fig.7 one can compare the resulting reduced $\chi^2$ values from a 
series of test runs where proton energy losses are not taken into account, as opposed to the same values 
from our actual simulations. An underestimation of the diffusion coefficient is evident in the 
simulations without losses. Furthermore, this underestimation explains the discrepancy between our 
results and those of B\"usching et al. \cite{Buschingetal2006}, as the diffusion coefficient calculated 
without losses is about two times smaller than when losses are taken into account in the propagation.

\begin{figure}
\includegraphics [width=0.48\textwidth]{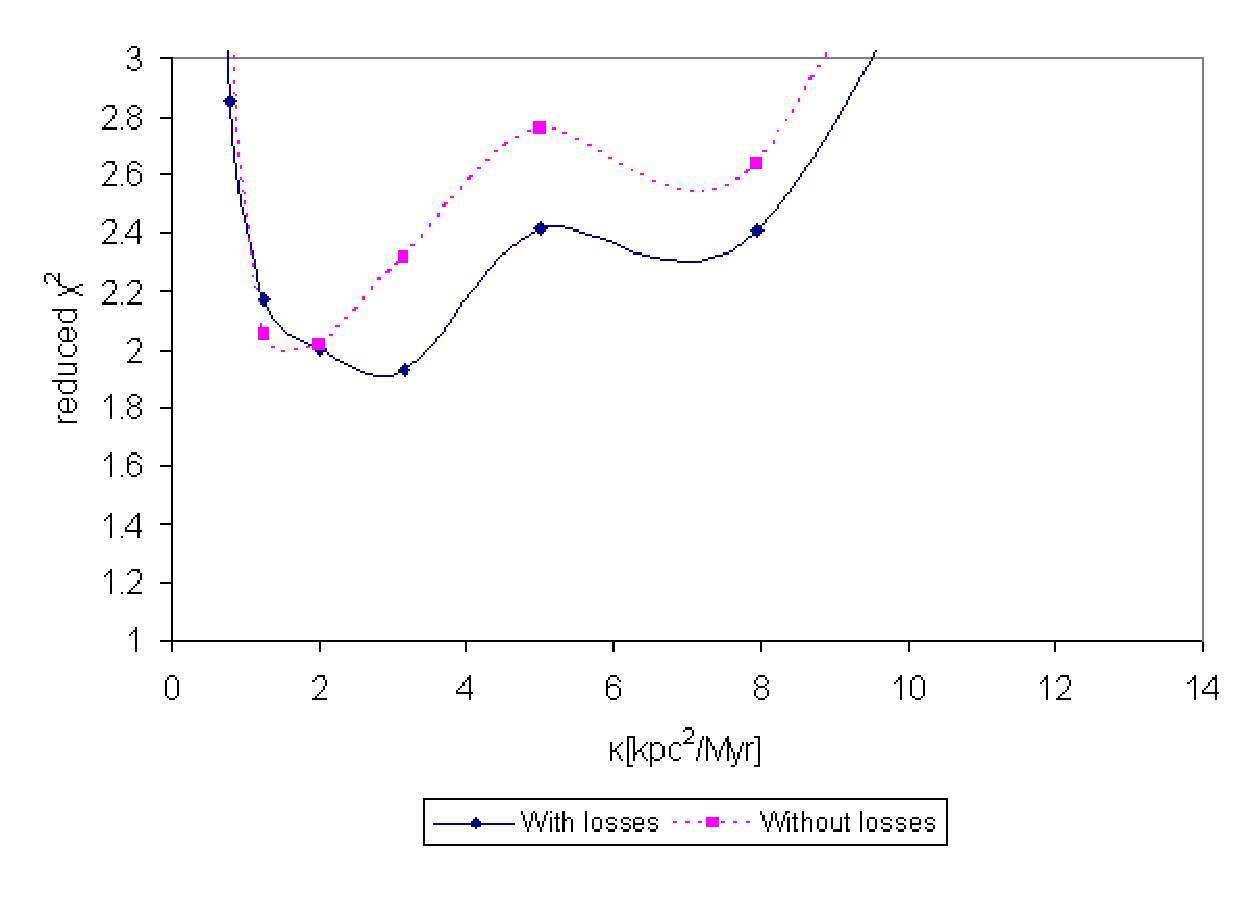}
\caption{Reduced $\chi^2$ values for different values of $\kappa_0$ and $\delta = 0.6$ for total propagation time $10^4 yr$ with and without proton energy losses taken into account during simulations.}\label{fig7}
\end{figure}

The above results were derived under the assumption that the acceleration of the cosmic rays responsible 
for the observed $\gamma$-rays occurred $10^4$ or $10^5$ yrs ago at a single source, with no subsequent 
periods of activity at that source. Recent papers \cite{ErlykinWolfendale2007} have noted that this may 
be a very simplified approach, as there have been many supernovae in the Galactic Centre region in the 
past millennia. Also, the uncertainty in the radial distances of hydrogen concentrations may have had a 
significant impact on the final results. Finally, these simulations assumed that the diffusion 
coefficient remains constant throughout the whole region of propagation, and local orderings of magnetic 
fields (including their correlation with molecular density within the gas clouds) were not taken into 
account.

Those limitations notwithstanding, the results of these simulations are useful in providing an estimate 
of the diffusion coefficient in the Galactic Centre, taking the different magnetic turbulence theories 
(that become manifest in the different values of $\delta$) into account. 

\section{Acknowledgments}
This project is co-funded by the European Social Fund and National Resources – (EPEAEK II) PYTHAGORAS II.

We would like to thank the anonymous referee for the useful comments which helped us improve this paper.

We would also like to thank John Kirk for helpful discussion.

\end{document}